\documentclass[preprint,aps,prb]{revtex4}

\usepackage{graphicx}

\usepackage{amsmath}

\begin{document}

\title{Geodesics and geodesic deviation in a two-dimensional
black hole}
\author{Ratna Koley}
\email{ratna@cts.iitkgp.ernet.in}
\author{Supratik Pal}
\email{supratik@cts.iitkgp.ernet.in}
\author{Sayan Kar}
\email{sayan@cts.iitkgp.ernet.in}

\affiliation{Department of Physics and Meteorology and
Centre for Theoretical Studies, Indian Institute of
Technology, Kharagpur 721 302, India}

\begin{abstract}
We introduce an exactly solvable example of timelike geodesic
motion and geodesic deviation in the background geometry of a
well-known two-dimensional black hole spacetime. The effective
potential for geodesic motion turns out to be either a harmonic
oscillator or an inverted harmonic oscillator or a linear
function of the spatial variable. corresponding to the
three different domains of a constant of the motion. The geodesic
deviation equation also is exactly solvable. The corresponding
deviation vector is obtained and the nature of the deviation is
briefly discussed by highlighting a specific case.
\end{abstract}

\maketitle

\section{Introduction}
Learning to solve the geodesic and geodesic deviation equations
is an integral part of a first course on general
relativity. However, most texts on general relativity\cite{rdi}
do not contain a sufficient number of solvable examples that
illustrate the behavior of geodesics in a nontrivial geometry.
This lack is largely due to the fact that the geodesic equations as
well as the geodesic deviation equations are not easy to solve
exactly in most of the standard geometries of interest. A
traditional example is the case of the line element on the surface of a
two-dimensional sphere for which
the geodesic and the geodesic deviation
equations can be solved analytically. Some other exactly solvable
examples in two dimensions include Rindler spacetime, the line
element on a cylinder and the hyperbolic plane.

In this article, we will discuss a two-dimensional example (with a
diagonal line element of Lorentzian signature ($-$$+$), i.e. of the form
 $ds^2=g_{00}dt^2
+ g_{11}dx^2$, with $g_{00}$ everywhere negative and $g_{11}$, positive)
, for which, like the sphere, there are exact
solutions for geodesics as well as geodesic deviation. The line
element for this example arises in the context of the stringy
black holes.\cite{string} In fact, this is the geometry of the
first stringy black hole discovered in the early nineties by
Mandal et al.\cite{msw} We will not go into the details of how this
line element is obtained. Interested readers may read the original
references\cite{msw} to enlighten themselves on this aspect. For
our purposes here, the line element represents a geometry that is
given to us, and we see that there are exact solutions for
geodesics and geodesic deviation for test particles moving in this
geometry.

\section{The Line Element}
We begin by writing down the line element of interest\cite{msw}: 
\begin{equation}
\label{1}
ds^{2} = -\bigl (1-\frac{M}{r} \bigr) dt^{2} +
\frac{kdr^{2}}{4r^{2}
\bigl (1-\frac{M}{r} \bigr)}.
\end{equation}
where M is equivalent to $GM/c^2$ with the choice of units $G=1$ and
$c=1$. Similarly $k$ is required to have dimensions of length squared
in order to render the metric coefficient $g_{rr}$ dimensionless.

The domain for the radial coordinate is $r>M$ and $t$ is allowed
to vary from $-\infty <t< \infty$. The physical meaning of the
constants $M$ and $k$ can be explained in the context of
two-dimensional string theory. For completeness (and also with the
risk of introducing some unexplained jargon), we mention that $M$
is related to mass and $k$ to the central charge parameter. 

Several purely geometrical facts about this line element are worth
noting. At $r=M$, $g_{tt} \rightarrow 0$ and
$g_{rr}\rightarrow
\infty$. If $r<M$, the signs of $g_{tt}$ and $g_{rr}$ are
interchanged. Readers familiar with some general relativity (say,
the Schwarzschild solution) would call $r=M$ the horizon and
thereby, characterize the line element as a black hole provided
there exists a singularity (which is the case here, as we shall
see later) inside the horizon (that is, in the region $r<M$). As
the title of this article suggests, the geometry represents a
two-dimensional black hole. As is common knowledge, a black hole
is a region from where nothing can escape. If an object falls into
a black hole (that is, the object crosses the horizon), the object
ends up getting stretched and torn apart by the gravitational
forces within. This distortion is quantified through the
deviation of geodesics. The ultimate fate is collapse into a
singularity where the matter density is infinitely large, thereby
being a region of extreme curvature. We shall concern ourselves
with the $r> M$ region.

The pathology in the line element (that is, $g_{rr} \to \infty $ ) at
$r=M$ is entirely a coordinate artifact (a poor choice of
coordinates). In addition, it is worth noting, that unlike the
case for similar but four-dimensional line elements where the
horizon is a sphere or some other two-dimensional surface, here it
is a point and falling into a black hole is manifest in
crossing this point and entering the region of no return. The
spacetime here is undoubtedly very similar to the two-dimensional
$r$--$t$ section of the static, spherically symmetric
four-dimensional black holes (like the Schwarzschild metric). Many of the
features of black holes can be understood via this example. In
fact an extensive research literature exists on two-dimensional
black holes where interesting phenomena such as Hawking radiation
and information loss have been discussed (see
Ref.~\onlinecite{string} and references therein and
Ref.~\onlinecite{strominger}).

The line element in Eq.~(\ref{1}) also can be written in the
following form\cite{strominger}:
\begin{equation}
\label{2}
ds^2 = -\tanh^2 \bigl (\frac{Mx}{2 \sqrt k}\bigr) dt^2 + dx^2,
\end{equation}
where we have used the coordinate transformation
$x = \frac{\sqrt k}{M} \cosh^{-1}(\frac{2r}{M}-1) > 0$.
Notice that the singular nature of the metric function
$g_{11}$ at $r=M$ is no longer there in this form of the line element.
A third form of the line element is the Kruskal extension or the
maximal analytical extension done exactly in the same way as for the
Schwarzschild metric. (For details on how to construct
Kruskal coordinates see any standard text on general relativity.)
If we introduce the coordinates $u$ and $v$ defined by:
\begin{subequations}
\label{3}
\begin{eqnarray}
u= t -\frac{\sqrt{k}}{2} \ln \vert r-M \vert \\
v= t +\frac{\sqrt{k}}{2} \ln \vert r-M \vert ,
\end{eqnarray}
\end{subequations}
and $U=-\sqrt{k}\exp({-u/\sqrt{k}})$, $V=\sqrt{k}
\exp({v/\sqrt{k}})$, we can rewrite the line element as:
\begin{equation}
\label{4}
ds^2 = -\frac{dUdV}{\frac{M}{\sqrt{k}}-\frac{UV}{k}} .
\end{equation}

In Eq.~(\ref{3}) the coordinates $u$ and $v$ run from $-\infty$ to
$\infty$ and $-\infty < U <0$, $0<V<\infty$. It is possible to further
rewrite the line element using coordinates $2T=V+U$, $2X=V-U$. The Kruskal form
of the line element is an `extension' over the whole domain of $r$ and $t$ 
The
geometry of this Kruskal extension has features similar to that
of the Schwarzschild.

\section{Timelike geodesics}\label{sec3}
Our main goal is to find timelike geodesics (that is,
trajectories of test particles of nonzero rest mass) in the
line element given in Eq.~(\ref{1}).
The equations and constraint for timelike geodesics are given as:
\begin{subequations}
\label{5}
\begin{eqnarray}
\ddot x^{\mu} + \Gamma^{\mu}_{\nu\lambda} {\dot x^{\nu}} {\dot
x^{\lambda}}&=&0\\ g_{\mu\nu}{\dot x^{\mu}}\dot x^{\nu} &=& -1,
\end{eqnarray}
\end{subequations}
where $x^{\mu}$ are the coordinates (here $r$ and $t$), and $\mu$
takes on values 0 and 1 for $t$ and $r$ respectively.
Equation~(5b) is the timelike constraint on the trajectories

The differentiation (indicated by dots) in Eq.~(\ref{5}) is with
respect to the affine
parameter (proper time) $\lambda$. The Christofel symbol
$\Gamma^{\mu}_{\nu\lambda}$ is given as:
\begin{equation}
\Gamma^{\mu}_{\nu\lambda}
= \frac{1}{2}g^{\mu\rho} [ g_{\rho\nu,\lambda}
+g_{\rho\lambda,\nu}- g_{\nu\lambda,\rho} ],
\end{equation}
where a comma denotes ordinary differentiation with respect to
coordinate $x^{\mu}$.The nonzero components of
$\Gamma^{\mu}_{\nu\lambda}$ for the line
element in Eq.~(\ref{1}) are
\begin{subequations}
\begin{eqnarray}
\Gamma^{0}_{10}&=&\Gamma^{0}_{01}=\frac{M}{2r(r-M)}\\
\Gamma^{1}_{00}&=&\frac{2M(r-M)}{kr}\\
\Gamma^{1}_{11}&=&-\frac{(2r-M)}{2r(r-M)}
\end{eqnarray}
\end{subequations}

The reader might ask why aren't we interested in null geodesics (that is, trajectories
of test particles having zero rest mass, such as light rays) ? In
two dimensions null geodesics are the same as those for the flat Minkowski line
element. Note that any two-dimensional line element (with Euclidean or
Lorentzian signature) can be written as $ds^2 =
\Omega^2(x,t) [ -dt^2 + dx^2]$, where $\Omega^2(x,t)$ is a
nonzero, positive definite function of $x$ and $t$ and is called
the conformal factor. Null geodesics require that
$g_{\mu\nu}{\dot x^{\mu}}\dot x^{\nu}=0$, which implies that in
two dimensions, null geodesics for Minkowski spacetime are the
same as those for any other nontrivially curved spacetime.

The timelike constraint in Eq.~(\ref{5}) for the line element of
interest here (Eq.~(\ref{1})), is given as:
\begin{equation}
\label{9}
-\bigl (1-\frac{M}{r} \bigr) {\dot t}^{2} + \frac{k}{4r^{2}
\bigl (1-\frac{M}{r} \bigr)} {\dot r}^{2} = -1.
\end{equation}
The geodesic equations (one each for $r$ and $t$) are
\begin{equation}
\label{10}
\dot t = \frac{E}{2\bigl (1-\frac{M}{r} \bigr)}
\end{equation}
\begin{equation}
\label{11}
\ddot r -\frac{1}{2} \bigl [ \frac{2r-M}{r^{2}-Mr} \bigr ] {\dot
r}^{2} + \frac{E^{2} Mr}{2k (r-M)} = 0,
\end{equation}
where in Eq.~(\ref{10}) we have substituted for $\dot t$ from
Eq.~(\ref{10}).
$E$ in the
above is a constant of the motion and arises due to the absence of the
coordinate $t$ in the metric coefficients. If we substitute
$\dot t$ from Eq.~(\ref{10}) in the
timelike constraint in Eq.~(\ref{9}). we obtain
\begin{equation}
\label{12}
{\dot r}^{2} = \frac{1}{k} [ (E^{2}-4) r^{2} + 4Mr ] = -V(r) +
E_{0},
\end{equation}
where $E_0$ is a constant and $V(r)$ is like an effective potential
given by:
\begin{equation}
V(r) = E_{0} - \frac{1}{k} [ (E^{2}-4) r^{2} + 4Mr ].
\end{equation}

Notice that the effective potential is like that of the harmonic
oscillator for $E^{2}<4$ and like that of an inverted
harmonic oscillator for $E^{2}>4$. For $E^{2} = 4$ it is linear with a
negative slope. We can make a coordinate change (translation) in
order to see the harmonic oscillator/inverted harmonic oscillator
forms explicitly. The potentials for the three cases are shown in
Fig.~1.

If we substitute $\dot r$ and its derivative with respect to
$\lambda$ from the timelike constraint (\ref{12}) in
Eq.~(\ref{11}) for $r$, the equation is
satisfied identically. Therefore, it is enough to just obtain
$r(\lambda)$ by integrating the timelike constraint for the three
cases ($E^{2}<4, E^{2}=4, E^{2}>4$) and obtain
$r(\lambda)$ and subsequently
$t(\lambda)$. These results are given below.

{\bf Case 1: $E^{2} <4$}.
\begin{subequations}
\begin{equation}
r(\lambda) = \frac{B}{2A} (1+ \cos a\lambda) \label{13}
\end{equation}
\begin{equation}
t(\lambda) = \frac{E}{2} \bigl [ \lambda + \frac{1}{qa}\ln
\frac{q+\tan c\lambda} {q-\tan c\lambda} \bigr ], \label{14}
\end{equation}
\end{subequations}
where $q^{2} =(1+f)/(1-f)$, $A=4-E^{2}$, $a=\sqrt{A/k}$,
$c=a/2$,
$f= 1-A/2$, and $B=4M$.

{\bf Case 2: $E^{2}=4$}.
\begin{subequations}
\begin{equation}
r(\lambda) = \frac{M}{k} \lambda^{2} \label{15}
\end{equation}
\begin{equation}
t(\lambda) = \lambda + \sqrt{\frac{k}{4}}\ln \bigl (\frac{\lambda -
\sqrt{k}} {\lambda + \sqrt{k}} \bigr) . \label{16}
\end{equation}
\end{subequations}

{\bf Case 3: $E^{2}>4$}.
\begin{subequations}
\begin{equation}
r(\lambda) =
\frac{B}{2A}
(-1+ \cosh
a\lambda) \label{17}
\end{equation}
\begin{equation}
t(\lambda) = \frac{E}{2} \bigl [ \lambda + \frac{q}{a}\ln \frac{\tanh
c\lambda -q}{\tanh c\lambda + q} \bigl ] ,
\end{equation}
\end{subequations}
where $q^{2} =(f-1)/(f+1)$,
$A=E^{2}-4$,
$a=\sqrt{A/k}$, $c=a/2$,
$f=1+A/2$, and $B =4M$.

In each case, the functional form of $r(t)$ can be obtained by utilizing the
parametric forms $r(\lambda)$, $t(\lambda)$. Figures~2--4
illustrate the above function. The origin of the coordinates
in each of these figures is chosen to be $r=1$ and $t=0$.
In particular, Fig.~4 shows the
actual radial trajectory $r(t)$ as a function of time. It is not
possible to write $r$ as an explicit function of $t$ because of the
nature of the functional dependence of $t$ on $\lambda$. However,
we may write $t(r)$ by
inverting
$r(\lambda)$ to $\lambda (r)$ and then using this function in
$t(\lambda(r))$. Notice that for $E^{2} < 4$, the trajectory hovers
around the horizon (extending only up to $r=2M=2$), reaching the value
$r=M=1$ only asymptotically. However, for $E^2 \ge 4$ there is no
such restriction and the trajectory can extend from $M$ to
$\infty$.

It should be mentioned that the ``discontinuities'' in some of the
plots arise because of the choice of the parameter ($M=1$ implies
that $r>1$) and, consequently, the
allowed domain of the independent variable. They are not real
discontinuities of the functions, as is apparent from the solutions.

\section{Geodesic deviation}
We now consider the geodesic deviation equation in order
to analyze the behavior of the spreading of the geodesic
curves obtained in Sec.~\ref{sec3}. Before we discuss the
geodesic deviation in the black hole line element, we
briefly recall some of the basic notions of geodesic
deviation.\cite{dev}

In a curved spacetime geodesics are not the straight lines of our
Euclidean intuition. For instance, the geodesics on the surface of
a sphere are the great circles. To understand how geodesics are
spread out in a spacetime of nontrivial curvature, we consider a
bundle of geodesics around a specific geodesic (called the central
geodesic). At every point on this central geodesic we erect little
normals. This normal is the connecting vector, so-called because
as we move along the connecting vector. we reach neighboring
geodesics. The reader might ask why normal deformations only? The
answer is that tangential deformations are taken into account
through the reparametrisation of the affine parameter $\lambda$,
or, in other words, they do not {\em deform} the trajectory to a
neighboring one.

It is obvious that the nature of the connecting vector (or
deviation vector) in a nontrivial spacetime depends crucially
on the properties of the spacetime, or more importantly, as it
turns out, on its curvature properties. For instance, on the
surface of a sphere, if we look at two neighboring great circle
arcs connecting the north and south poles, the distance between
them increases, reaches a certain value at the equator, and then
decreases untill it reaches a zero value at the other pole. The
deviation vector essentially measures this effect, which is
dictated by the properties of the Riemann tensor
$R^{\mu}_{\nu\lambda\rho}$ given by:
\begin{equation}
R^{\mu}_{\nu\lambda\rho} =
\Gamma^{\mu}_{\nu\rho,\lambda} -\Gamma^{\mu}_{\nu\lambda,\rho}
+\Gamma^{\mu}_{\sigma\lambda}
\Gamma^{\sigma}_{\nu\rho} - \Gamma^{\mu}_{\sigma
\rho}\Gamma^{\sigma}_{\nu
\lambda} .
\end{equation}
The Ricci tensor and the Ricci scalar are obtained from the above
Riemann tensor by appropriate contractions with the metric tensor (Ricci tensor,
$R_{\mu\nu} = R^{\lambda}_{\mu\lambda\nu}$ and Ricci scalar, $R = g^{\mu\nu}R_{\mu\nu}$).

We know from Einstein's general relativity that geometry is
equivalent to the presence of a gravitational field. Thus the
deviation of trajectories from each other is
a measure of the relative gravitational force between the objects
moving on these separate, neighboring, trajectories. We now state
the geodesic deviation equation and solve it for the geodesics in
the black hole line element, Eq.~(\ref{1}), discussed here.

Given a geodesic curve, we identify the tangent and normal to it as the
vectors $e^{\mu}$ and $n^{\mu}$ respectively. These satisfy the
orthonormality conditions:
\begin{subequations}
\label{20}
\begin{eqnarray}
g_{\mu\nu}e^{\mu}e^{\nu} &=& -1 \\
g_{\mu\nu}n^{\mu}n^{\nu} &=& 1 \\
g_{\mu\nu}e^{\mu}n^{\nu} &=& 0 .
\end{eqnarray}
\end{subequations}
We make the assumptions:
\begin{subequations}
\label{21}
\begin{eqnarray}
e^{\mu} &\equiv& (\dot t,\dot r ) \\
n^{\mu} &\equiv& (f(r)\dot r, g(r)\dot t ).
\end{eqnarray}
\end{subequations}
Note that other choices can be made and thus the above definitions in Eq.~(\ref{21})
are by no means unique. If we substitute the assumptions in Eq.~(\ref{21})
into the orthonormality conditions Eq.~(\ref{20}), we obtain the
following relations for
$f(r)$ and $g(r)$:
\begin{subequations}
\begin{eqnarray}
f(r) &=& \sqrt{\frac{g_{11}}{-g_{00}}} \\
g(r) &=& \sqrt{\frac{-g_{00}}{g_{11}}}.
\end{eqnarray}
\end{subequations}
Therefore the normal vector $n^\mu$ takes the form:
\begin{subequations}
\begin{eqnarray}
n^0 &=& \frac{\sqrt k}{2r(1- \frac{M}{r})} \dot r \\
n^1 &=& \frac{2r(1- \frac{M}{r})}{\sqrt k} \dot t.
\end{eqnarray}
\end{subequations}
The equation for the normal deformations $\eta$ (where $\eta^{\mu}
= \eta n^{\mu}$, $\eta^\mu$ being the deviation vector) is given
by:
\begin{equation}
\frac{d^{2}\eta}{d\lambda^{2}} + R_{\mu\nu\rho\sigma}e^{\mu}n^{\nu}e^{\rho}
n^{\sigma} \eta = 0.
\end{equation}
The nonzero components of Riemann tensor in the coordinate frame are
given by:
\begin{subequations}
\begin{eqnarray}
R_{0101} &=& R_{1010} = -\frac{M}{2r^3} \\
R_{0110} &=& R_{1001} = \frac{M}{2r^3}.
\end{eqnarray}
\end{subequations}

There is actually only one independent component of the Riemann
tensor in two dimensions (recall that in $n$ dimensions, the
number of independent components is $n^2(n^2-1)/12$).
The nonzero Ricci tensor components in the coordinate frame
($R_{\mu\nu} = R^{\alpha}_{\mu\alpha\nu}$) and the Ricci scalar
($R=g^{\mu\nu}R_{\mu\nu}$) are
\begin{subequations}
\begin{eqnarray}
R_{00}&=&-\frac{2M(r-M)}{kr^2} \\
R_{11}&=&\frac{M}{2r^2(r-M)}\\
R&=&\frac{4M}{kr}
\end{eqnarray}
\end{subequations}

The Riemann tensor components and Ricci scalar diverge as $r\rightarrow 0$.
The geometry becomes singular as $r\rightarrow 0$. However, $r\rightarrow M$
does not seem to be a preferred point (in comparison to $r=0$ say), a fact
that re-emphasizes the point that the pathology at $r=M$ in the line
element in Eq.~(\ref{1}) is largely a coordinate artifact. Of course,
the special nature of $r=M$ is that it is a horizon with the metric
signature changing sign as we move from $r>M$ to $r<M$.

If we substitute the Riemann tensor components for the two-dimensional 
black hole line element and use the normal and tangent mentioned in Eq.~\ref{21}
, we find that the deviation equation
reduces to the following equations for the three cases discussed
earlier.

{\bf Case 1: $E^{2}<4$}.
\begin{equation}
\label{27}
\frac{d^{2}\eta}{d {\bar \lambda}^{2}} -
2 \sec^2 {\bar \lambda} \eta = 0 ,
\end{equation}
where $\bar\lambda = \frac{1}{2}\sqrt{\frac{4-E^2}{k}}\lambda$.
The two linearly independent solutions to Eq.~(\ref{27}) are
\begin{equation}
\label{28}
\eta (\bar\lambda) = C_1 \tan {\bar\lambda} + C_2 ({\bar\lambda}
\tan {\bar\lambda} + 1).
\end{equation}

{\bf Case 2: $E^{2} = 4$}.
\begin{equation}
\label{29}
\frac{d^{2}\eta}{d {\lambda}^{2}} - \frac{2}{\lambda^2}\eta =0.
\end{equation}
The solutions are
\begin{equation}
\label{30}
\eta (\lambda) = C_{1} \lambda^2 + C_2 \frac{1}{\lambda}
\end{equation}

{\bf Case 3: $E^{2}>4$}.
\begin{equation}
\label{31}
\frac{d^{2}\eta}{d {\bar \lambda}^{2}} -2 {\rm
cosech}^{2}{\bar\lambda} \eta =0,
\end{equation}
where $\bar\lambda = \frac{1}{2}\sqrt{\frac{E^2-4}{k}}\lambda$.
In this case, the solutions are
\begin{equation}
\label{32}
\eta (\bar\lambda) = C_1 \coth {\bar\lambda} + C_2 (\bar\lambda
\coth \bar\lambda -1)
\end{equation}
In Eqs.~(\ref{28}), (\ref{30}), and (\ref{32}), $C_1$ and $C_2$ are
separate arbitrary constants in each case. These constants should
be determined by the initial conditions (that is, the value of the
initial separation between a pair of geodesics at a specific value
of the affine parameter and the value of the rate of change of the
initial separation at the same value of
$\lambda$, similar to
the conditions for obtaining particular solutions of second-order,
linear, ordinary differential equations).

Now that we have obtained $\eta$ for each case, it is easy to
write down the deviation vectors, which are given by
$\eta^{\mu} = \eta n^{\mu}$ (where $\mu = 0,1$). Let us now discuss a
representative case, say $E^2=4$. The range of $\lambda$ (for $r>M
=1$) is between $-\infty <\lambda <-1$ and
$1<\lambda <\infty$. If we consider a pair of neighboring geodesics at
say $\lambda =1.5$ and assume $\dot \eta$ is positive, then we are
forced to choose the solution for $\eta$ to be proportional to
$\lambda^2$. The evolution of $\eta$ then suggests that as we move
further and further to larger positive $\lambda$, the geodesics
spread out (diverge). On the other hand, if we assume $\dot \eta$
to be negative, then we must choose the $\frac{1}{\lambda}$
solution, which says that for larger
$\lambda$ (which corresponds to larger $r$ as well), the geodesics
converge, at least locally. However, as we go to larger
$\lambda$, $\dot \eta$ also becomes smaller and approaches
zero as $\lambda \rightarrow \infty$ ($r\rightarrow \infty$).
Therefore, ultimately the geodesics become parallel to each other.

It is also useful to note that $d^2 \eta/d \lambda^2$, the measure of the
gravitational force, is like a
relative force  which acts to change the separation between
a pair of neighboring geodesics. We can easily substitute the values of
$\lambda$ (or $\bar\lambda$) from the expressions
of $r(\lambda)$ in the geodesic deviation equations~(\ref{27}), (\ref{29}), and (\ref{31}) to analyze the nature of the force. As an example,
let us consider the case
$E^2=4$.
For $\eta = \lambda^2$, the relative force is constant throughout and is
given by:
\begin{equation}
\frac{d^2 \eta}{d \lambda^2} = 2,
\end{equation}
whereas for $\eta = 1/ \lambda$, we have
\begin{equation}
\label{temp}
\frac{d^2 \eta}{d \lambda^2} = 2 \biggl(\frac{M}{kr} \biggr)^{3/2} .
\end{equation}
Equation~(\ref{temp}) tells us that the force vanishes as $r$ goes to
infinity (recall that the geometry is flat in this limit
and the geodesics will become parallel to each other),
whereas for $r \to M$, it has the value $2/k^{3/2}$. A similar analysis can be carried out for the
cases with $E^2 <4$ and $E^2>4$.

A point worth mentioning here (without proof) is that in the
above geometry, the geodesics do not seem to converge to a point
within a finite value of $\lambda$ in any of the cases because of the 
fact that the Ricci scalar is non-negative
(it is zero only when $r\rightarrow
\infty$). In two dimensions, timelike
geodesics do not converge to a point (focus) within a finite value
of the affine parameter unless $R \le 0$.\cite{sk}

\section{Conclusions}
In summary, the exact solutions of the geodesic deviation equation
can provide us
with a better understanding of the nature of the separation
between geodesics.
Readers can further investigate the deviation vectors in the cases
not discussed above in order
to improve their understanding of geodesic deviation. Trajectories in
two dimensions are of the form $r(t)$. The nontrivial curvature of the spacetime
is responsible for their difference from the usual `straight lines' in 
Minkowski spacetime--the solutions to the geodesic equation 
obtained in the above line element provide examples. 
If we take pairs of trajectories it is obvious that
they may converge towards
or diverge away from each other--a measure of this effect is geodesic deviation.
We have illustrated this effect in our example above.   

Our aim has been to provide an example
where the geodesic and
geodesic deviation equations can be {\em exactly} solved. It has been shown
that for the line element of a two-dimensional black hole, exact,
reasonably simple
solutions do exist. We hope that the relevance of the line element,
as well as the simplicity of the solutions will attract students
and teachers to use this example when teaching a first course in
general relativity. We also mention that, as far as we know, the
exact solutions to the geodesic and the geodesic deviation 
equations do not exist
in the literature and might be of some interest to researchers as
well. There are innumerable solutions representing black holes and
cosmologies in two-dimensional gravity, most of which have
appeared largely in the last decade and a half. We believe that a carefully
chosen section of these line elements can be used as worthwhile
teaching tools in a first course on general relativity. 

\begin{acknowledgements}
SP thanks the Centre for Theoretical Studies, IIT Kharagpur, India for
a visit during which this work was done. RK acknowledges financial
support from IIT Kharagpur, India. 
\end{acknowledgements}

\newpage

\section*{Figure Captions}

\vspace{.3in}

{\bf Fig.~1:} Effective potential ($y$-axis) as a function of $r$
($x$-axis) for three cases $E^2 = 1$ (Fig. $1a$), 4 (Fig. $1b$), 7 (Fig. $1c$)
respectively; $E_0 = 10$, $k = 1$, $M = 1$ for all three cases.

{\bf Fig.~2:} $r(\lambda)$ ($y$-axis) versus $\lambda$ ($x$-axis)
as in Eqs.~$13a$ (Fig. $2a$), $14a$ (Fig. $2b$) and $15a$
(Fig. $2c$) with $A=3$, $k=1$, and $M=1$.

{\bf Fig.~3:} $t(\lambda)$ ($y$-axis) versus $\lambda$ (x-axis) as
in Eqs.~$13b$ (Fig. $3a$), $14b$ (Fig. $3b$) and
$15b$ (Fig. $3c$) with $A=3$, $k=1$, and $M=1$.

{\bf Fig.~4:} $r(t)$ ($y$-axis) versus $t$ ($x$-axis) from Eqs.~$13a$ and $13b$ 
(Fig. $4a$), $14a$ and $14b$ (Fig. $4b$), $15a$ and $15b$ (Fig. $4c$) with
$A=3$, $k=1$, and $M=1$.

\newpage

\begin{figure}

\includegraphics[width=5cm,height=4cm]{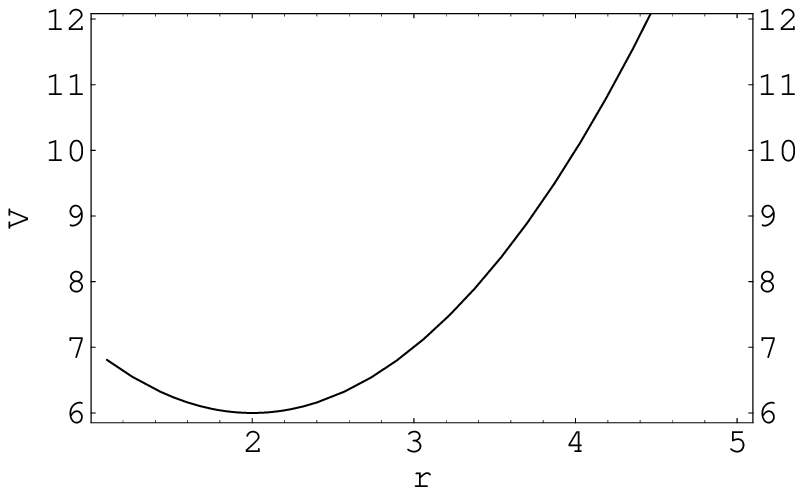}\\

\hspace{.5cm}Fig.$1a$: $E^2<4$
\vspace{1cm}

\includegraphics[width=5.7cm,height=4.5cm]{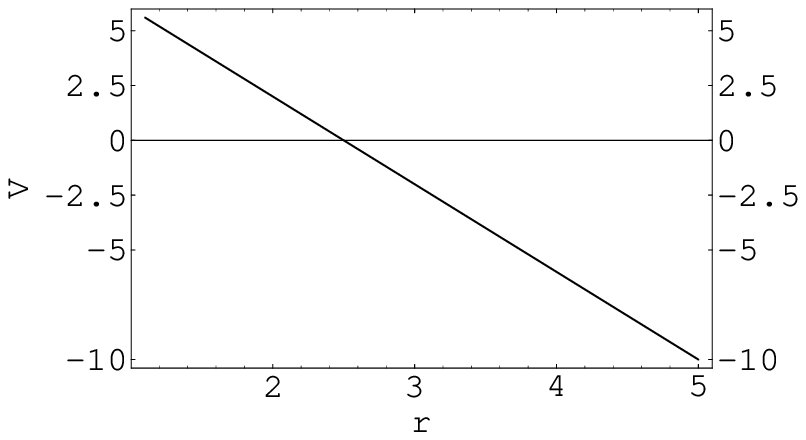}\\

\hspace{.5cm}Fig.$1b$: $E^2=4$ 
\vspace{1cm}

\includegraphics[width=5.7cm,height=4.5cm]{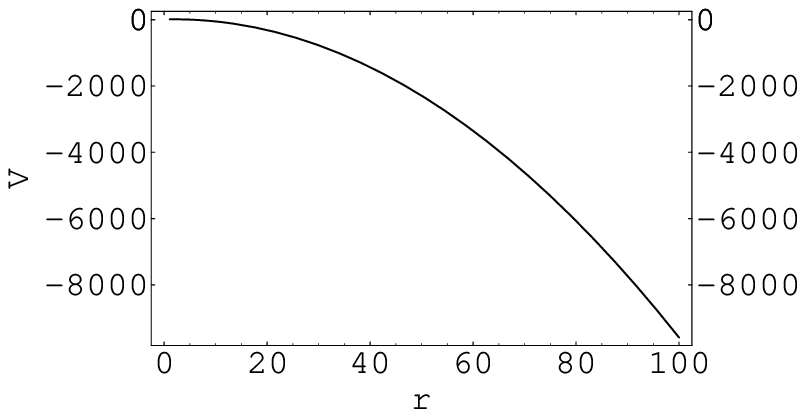}\\
\hspace{.5cm} Fig.$1c$: $E^2>4$
\vspace{1cm}

\noindent {\bf Fig.1}: Effective potential ($y$-axis) as a function of $r$
($x$-axis) for three cases $E^2 = 1$ (Fig. $1a$), 4 (Fig. $1b$), 7 (Fig. $1c$)
respectively; $E_0 = 10$, $k = 1$, $M = 1$ for all three cases.\\

\end{figure}

\newpage

\begin{figure}

\includegraphics[width=5cm,height=4.5cm]{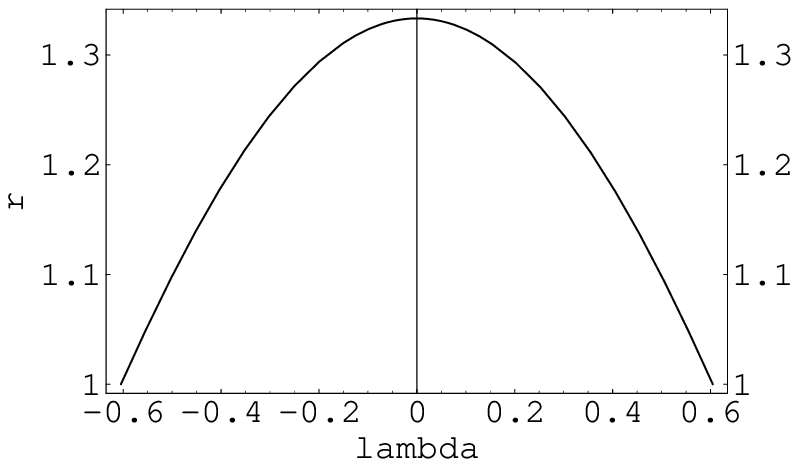}\\

\hspace{.5cm} Fig. $2a$: $E^2<4$
\vspace{1cm}

\includegraphics[width=5cm,height=4.5cm]{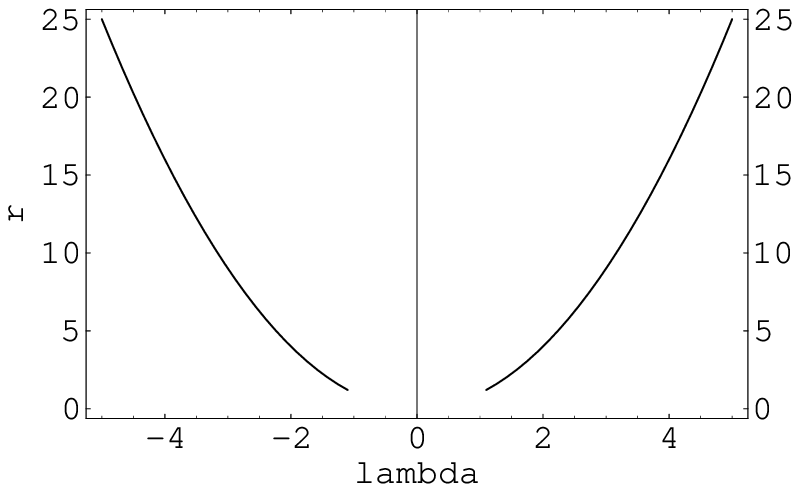}\\

\hspace{.5cm} Fig. $2b$: $E^2=4$
\vspace{1cm}

\includegraphics[width=6cm,height=4.5cm]{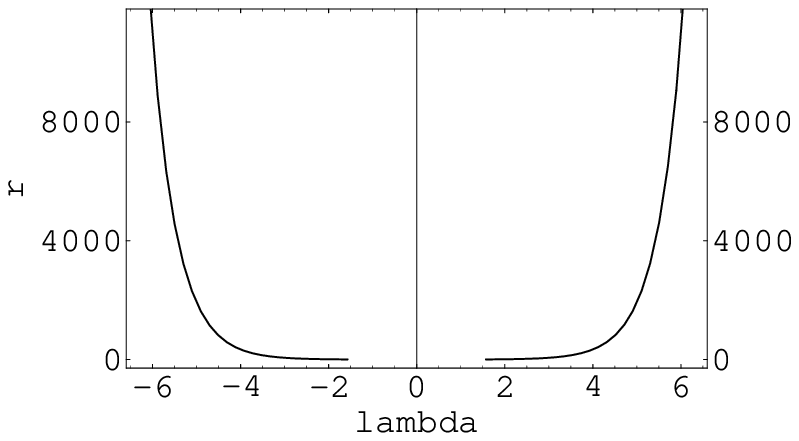}\\

\hspace{.5cm} Fig. $2c$: $E^2>4$
\vspace{1cm}

\noindent {\bf Fig.2}: $r(\lambda)$ ($y$-axis) versus $\lambda$ ($x$-axis)
as in Eqs.~$13a$ (Fig. $2a$), $14a$ (Fig. $2b$) and $15a$
(Fig. $2c$) with $A=3$, $k=1$, and $M=1$.\\
\end{figure}

\newpage
\begin{figure}

\includegraphics[width=5cm,height=4cm]{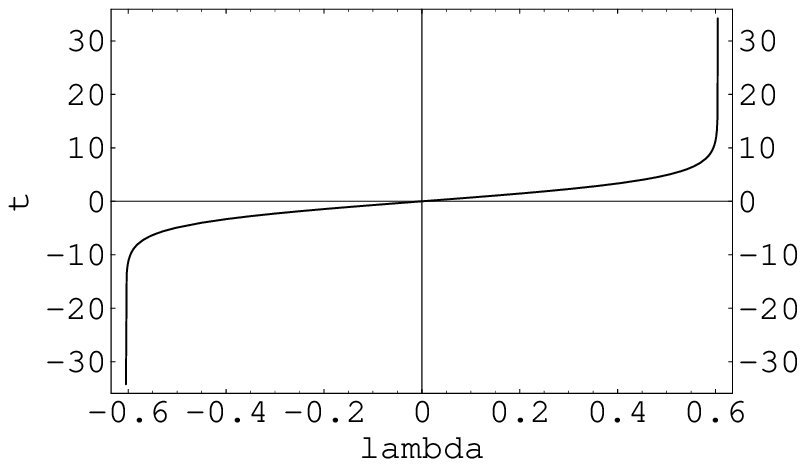}\\

\hspace{.5cm} Fig. $3a$: $E^2<4$
\vspace{1cm}

\includegraphics[width=5cm,height=4cm]{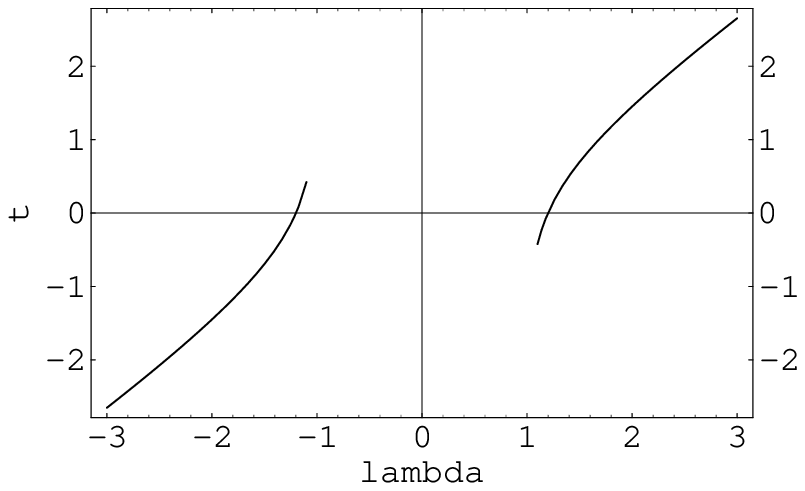}\\

\hspace{.5cm} Fig. $3b$: $E^2=4$
\vspace{1cm}

\includegraphics[width=5cm,height=4cm]{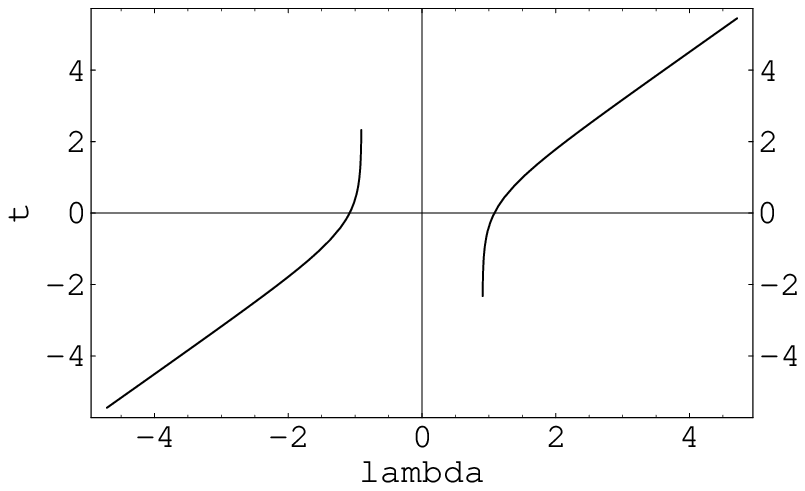}\\

\hspace{.5cm} Fig. $3c$: $E^2>4$
\vspace{1cm}

\noindent {\bf Fig.3}: $t(\lambda)$ ($y$-axis) versus $\lambda$ (x-axis) as
in Eqs.~$13b$ (Fig. $3a$), $14b$ (Fig. $3b$) and
$15b$ (Fig. $3c$) with $A=3$, $k=1$, and $M=1$.\\
\end{figure}

\newpage
\begin{figure}

\includegraphics[width=5cm,height=4.5cm]{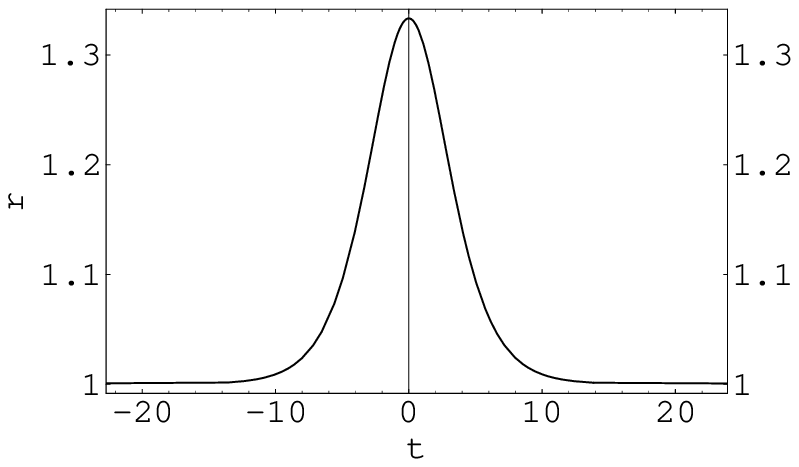}\\

\hspace{.5cm} Fig. $4a$: $E^2<4$
\vspace{1cm}

\includegraphics[width=5cm,height=4.5 cm]{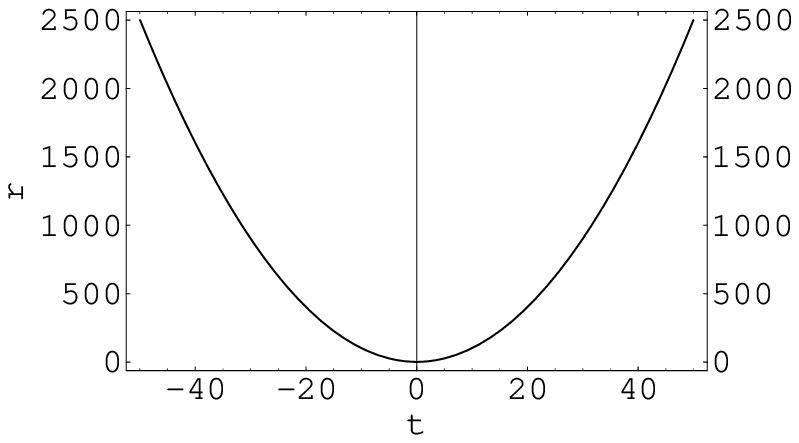}\\

\hspace{.5cm} Fig. $4b$: $E^2=4$
\vspace{1cm}

\includegraphics[width=5cm,height=4.5cm]{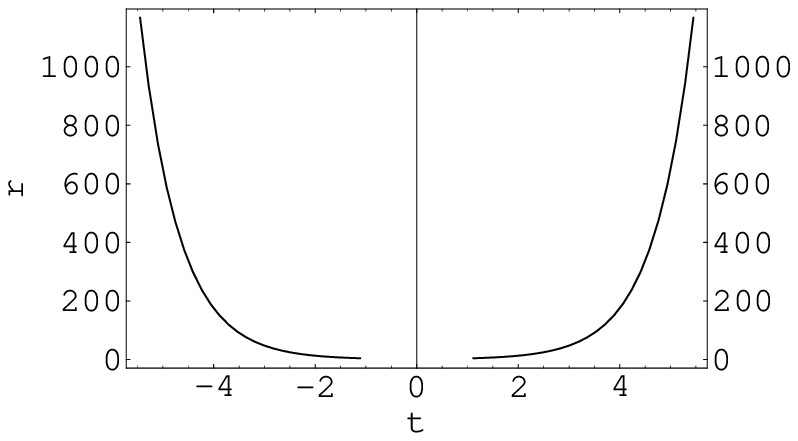}\\
  
\hspace{.5cm} Fig. $4c$: $E^2>4$
\vspace{1cm}

\noindent {\bf Fig.4}: $r(t)$ ($y$-axis) versus $t$ ($x$-axis) from Eqs.~$13a$ and $13b$ 
(Fig. $4a$), $14a$ and $14b$ (Fig. $4b$), $15a$ and $15b$  (Fig. $4c$) with $A=3$, $k=1$, and 
$M=1$ in all three cases.\\
\end{figure}


\begin{thebibliography}{99}
\bibitem{dev} A good introduction to geodesics and geodesic deviation
in particular is available in the text by R. D'Inverno refered to
in Ref.~\onlinecite{rdi}.
\end{thebibliography}
\end{document}